# Speech Enhancement via Deep Spectrum Image Translation Network

Hamidreza Baradaran Kashani
Electrical Engineering Faculty
Amirkabir University of Technology
Tehran, Iran
hr_baradaran@aut.ac.ir

Ata Jodeiri
School of Electrical & Computer Engineering
College of Engineering, University of Tehran
Tehran, Iran
ata.jodeiri@ut.ac.ir

Mohammad Mohsen Goodarzi
Department of Electrical and Computer Engineering,
Buein Zahra Technical University,
Qazvin, Iran
mm.goodarzi@bzte.ac.ir

Iman Sarraf Rezaei
Speech and Language Processing Group
Research Center for Development of Advanced Technologies
Tehran, Iran
sarraf@rcdat.ir

*Abstract*—Quality and intelligibility of speech signals are degraded under additive background noise which is a critical problem for hearing aid and cochlear implant users. Motivated to address this problem, we propose a novel speech enhancement approach using a deep spectrum image translation network. To this end, we suggest a new architecture, called VGG19-UNet, where a deep fully convolutional network known as VGG19 is embedded at the encoder part of an image-to-image translation network, i.e. U-Net. Moreover, we propose a perceptually-modified version of the spectrum image that is represented in Mel frequency and power-law non-linearity amplitude domains, representing good approximations of human auditory perception model. By conducting experiments on a real challenge in speech enhancement, i.e. unseen noise environments, we show that the proposed approach outperforms other enhancement methods in terms of both quality and intelligibility measures, represented by PESQ and ESTOI, respectively.

*Keywords—speech enhancement; hearing aid; image-to-image translation; VGG19-UNet; perceptually-modified spectrum image.*

## I. Introduction

Recorded human speech is usually subject to distortions caused by reverberation and background noise. These distortions degrade speech intelligibility and speech quality and have negative effects on performance of many real applications of speech technology, such as automatic speech recognition (ASR) [1] and speaker identification (SI) [2]. Much more importantly, speech enhancement under additive background noise is a vital requirement for hearing aid and cochlear implant users in daily life [3].

A lot of single-channel speech enhancement methods are proposed and could be classified in following categories:

1- *Spectral estimation methods*: focused on exploring the statistical features of speech and noise spectrums and their differences. The most famous and still powerful method of this category is so called OMLSA [4] which is based on an improved minimum mean-square error estimation (MMSE) of log-spectral of clean speech. Most of these methods are unsupervised and try to estimate noise spectrum using the additive nature of the background noise, or the statistical properties of the speech and noise signals. Then extract it from spectrum of noisy signal. These assumptions make this category vulnerable to non-stationary noise in real-world scenarios and unexpected acoustic conditions.

2- *Source separation methods*: In these methods, the noisy speech is assumed to be the mixture of multiple sources contacting original speech signal plus noise and other degrading sources. Then the process of speech enhancement is formed as a source separation method.

3- *Mapping methods*: These methods try to map noisy spectrum to clean one. Due to the great ability of Neural Network (NN) in forming a non-linear mapping, and their fast development, NNs are the most adopted method in this category. To cover spectral space of speech signal and divers types of noise, recently Deep Neural Networks (DNN) attracted more attentions.

DNNs are employed in various structures to map noisy speech to clean one. In [5], a DNN called Deep Auto Encoder (DAE) was trained to map clean speech data to itself. It is claimed that when this DAE structure is fed with noisy speech, it will produce clean counterpart. In [6] the same DAE structure is used to map paired noisy-clean speech to clean speech and some improvements reported using this trick. It is also shown that DNN used for speech enhancement produces less musical artifact [7] compared to spectral estimation methods. Although DNN based enhancement methods suffer from unseen noise problem, the vast DNNs capacity allows them to be trained on wide variety of noise types during training process [8]. This big capacity allows them to be used for both de-reverberation and de-noising simultaneously without necessary change in DNN topology [9].

To address the problem of unseen noise and the harder situation of mixed types of noise, [10] proposed to use environment specific noises (such as office noises) and take advantage of psychoacoustic models in a noise aware training procedure. Good results of using psycho acoustic metrics attracted more attentions [11], [12]. While most DNN based speech enhancement methods use mean square error (MSE) as loss function in DNN training phase, Martin et al. [12] proposed a perceptual metric for speech quality as loss function and claimed significant improvements. Although fully connected feedforward DNN (FCDNN) structures achieved good results, but the huge number of parameters makes them expensive. On the other hand, the weight tying mechanism of Convolutional Neural Network (CNN) will result in less parameters compared to equivalent FCDNN. This makes CNN attractive for low resources situations such as speech enhancement in hearing aids [13] and medical image processing [14]. On the other hand, the inherent feature extraction property of CNN encouraged researchers to implement time domain and waveform end-to-end speech enhancement networks [15], [16].

In many real world applications such as mobile communication, hearing aids and cochlear implants, it is necessary to have very low latency. Tan et al. [17], [18] combined CNN with Long Short Term Memory (LSTM) Recurrent Neural Network (RNN) to achieve this goal. Some studies tried to combine DNN mapping methods with spectral estimation methods such as MMSE estimators [19], [20]. More complex networks such as Generative Adversarial Networks (GANs) are also used for speech enhancement [21] and reasonable results achieved in unseen noise conditions.

In this paper, we specifically focus on enhancing speech signals under additive background noises by exploiting a fully convolutional network (FCN) instead of fully connected feedforward networks, which do not fully utilize the spectral structures of the speech. As a basis, we first consider one of the well-known types of FCNs, namely U-Net [22], presented as an image-to-image translation network. We then propose a novel architecture based on U-Net, called *VGG19-UNet*, where a deep fully convolutional network known as VGG19 [23] is embedded at the encoder part of the encoder-decoder network architecture, i.e. U-Net. Notably in our previous research [24], we employed U-Net for the task of speech declipping and achieved promising results.

In order to learn the proposed network, the input and output features are considered as 256x256 spectrum images extracted from noisy and clean speech signals, respectively. However, we propose to employ spectrum images in Mel frequency and power-law non-linearity amplitude domains, representing good approximations of human auditory model, instead of conventionally linear frequency and logarithm amplitude domains used, respectively.

The rest of the paper is organized as follows: Section II firstly describes the architectures of U-Net, VGG and the proposed one, i.e. *VGG19-UNet*. Then, we bring forward the proposed speech enhancement approach. Section III details the experimental results of the proposed approach in comparison with other well-known speech enhancement methods. Finally, conclusions are presented in Section IV.

II. METHOD

*A. U-Net*

U-Net is one of the well-known convolutional networks for the fast and precise image-to-image translation tasks such as image segmentation and image de-noising. It consists of three general parts, which are encoder, decoder, and skip connections. U-Net passes the feature maps from each level of the encoder over to a similar level in the decoder. In the encoder, local and structural features are extracted. The input size of the image is reduced through the encoder path in order to increase the receptive field, make the model robust to noise and artifact, and also decrease the computational cost. Increasing the receptive field leads to propagating global information in both time and frequency domains. The skip connections allow the U-Net to consider features at various scales by combining the local and global feature maps. In fact, encoder path captures the context of the input image, and decoder path extracts the abstract features. Moreover, the precise localization between symmetric feature maps is done via skip connections.

The capability of the U-Net in learning from a relatively small dataset made it a suitable choice in dealing with medical images, whereas manual preparation of the masks is a very costly procedure. Typically using a pre-train model for avoiding the over-fitting issue in the training of the deep convolutional networks seems necessary. In this case, firstly the network weights are initialized on relatively large, millions of images such as ImageNet public dataset and then are transferred to the target dataset. Several studies have shown that U-Net can easily train from scratch and converge very fast without the need for pre-train model.

*B. VGG Architecture*

The main contribution of the VGG was to reveal the impact of the convolutional network depth on the performance of the network in large-scale image localization and classifications tasks.

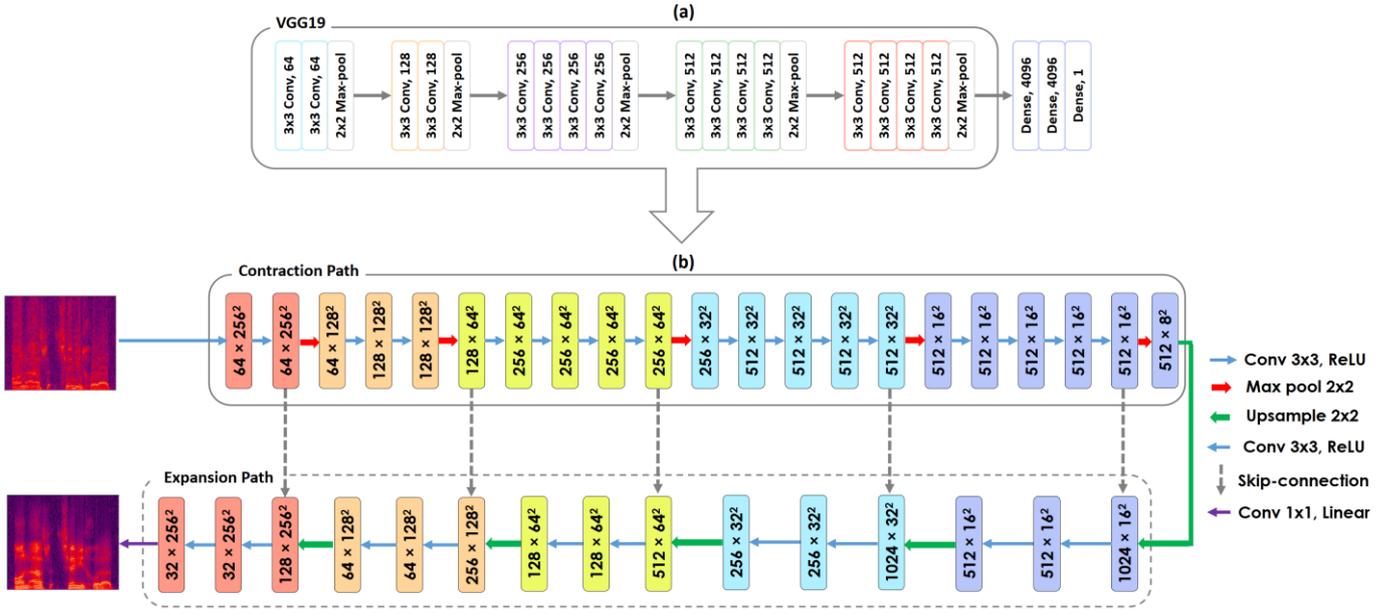

Figure 1. Representations of (a) VGG19 architecture and (b) the proposed architecture: VGG19-UNet

The networks prior to VGG used a large kernel size such as 7×7 and 11×11 to increase the receptive field, but VGG showed that a large receptive field that is achieved by increasing the depth of the convolutional layers with 3×3 kernel size leads to significantly higher accuracy.

As shown in Figure 1(a), VGG19 consists of 16 3×3 CNN layers in five convolutional blocks for feature extraction and three fully connected layers for classification. The first CNN produces 64 channels and then, at each convolutional block as the network deepens, the number of channels doubles until it reaches 512. All CNN layers are equipped with the Rectified Linear Unit (ReLU) non-linearity. Additionally, spatial pooling is carried out by five 2×2 max-pooling layers, such that each pooling follows one convolutional block and reduces the size of the input feature map.

### C. Proposed Architecture: VGG19-UNet

In some studies, by employing VGG instead of the previous generation of the models, it has been demonstrated that the representation depth is beneficial for improving the performance of the system. Motivated by that, we propose a novel encoder-decoder style network for speech enhancement task. As shown in Figure 1(b), to construct the encoder part, we use the five convolutional blocks of the VGG19 as a powerful deep feature extractor. The latest extracted features are fed to the decoder part. In the decoder, a 2×2 up-sampling and two 3×3 CNN layers with ReLU non-linearity are applied and then this sequence is repeated five times. Contrasting the encoder path, the number of channels in each sequence is halved, eventually reaches 32 and then is converted to the spectrum images via one 1×1 CNN with linear activation function. It is notable that, after each up-sample layer, the same-size corresponding tensors in the encoder and decoder parts are concatenated.

### D. Proposed Approach

The proposed approach, namely *VGG19-UNet-MelPow* speech enhancer, consists of three steps as follows:

- Step 1) Extracting perceptually-modified spectrum images

Firstly, we apply short-time Fourier transform (STFT) to the (noisy/clean) speech signals and compute the short-time feature vectors representing the magnitude spectrum. To compute STFT, the frame length, frame shift and the number of Fourier transform points are selected as 32ms, 8ms and 512, respectively. As a result, the dimension of each magnitude spectrum vector is 257.

According to the experiences of our studies presented in [25], [26], we found that using Mel frequency, especially by direct warping of the spectrum to Mel scale, followed by power-law non-linearity with exponent (2/15) generates more robust speech features than employing linear frequency and logarithm non-linearity. In this study, therefore, we sequentially apply the Mel warping and power-law non-linearity with exponent (2/15) to the magnitude spectrum vectors to make new vectors called *MelPow* magnitude spectrum vectors. In the following, perceptually-modified magnitude spectrum images are made by concatenating 256 successive *MelPow* magnitude spectrum vectors. It is notable that in order to keep the image symmetry, the highest frequency bin is ignored. Consequently, each spectrum image is a 256x256 image representing the magnitude values at 256x256 time-frequency units. Note that before making spectrum images, only for noisy speech, mean-variance normalization is applied to the *MelPow* magnitude spectrum vectors at the speech utterance level. At the end of this step, for either training or development set, an image dataset, consisting of perceptually-modified magnitude spectrum images of noisy signals and the corresponding clean ones, is provided.

- Step 2) Learning VGG19-UNet architecture

The set of training spectrum images provided from the previous step is employed to learn the VGG19-UNet network. By using Tensorflow and Keras library, the proposed network is optimized by minimizing Adam algorithm with 0.0002 learning rate for 50 epochs. At each epoch, a mini-batch size of 10 images is randomly selected from the training images and fed into the network to learn the parameters. We use the linear activation function to generate the enhanced spectrum image. Moreover, the mean square error (MSE) as the cost function is minimized during training. Finally, the learned model resulting in the minimum of MSE on the development set is selected for generating the enhanced speech signals.

- Step 3) Generating enhanced speech

At the enhancement step, firstly, the perceptually-modified magnitude spectrum images of the input noisy speech are extracted as stated at Step 1. The input images are fed to the learned VGG19-UNet model to produce the enhanced spectrum images in Mel frequency and power-law amplitude domains. In the following, the inverse mappings to linear amplitude and linear frequency domains are sequentially applied to the spectrum images. Finally, the phase information of the noisy signal is directly combined with the enhanced magnitude spectrum and inverse STFT is then used to generate the time-domain enhanced speech.

### III. EXPERIMENTAL RESULTS

#### A. Dataset Preparation

FARSDAT corpus [27] was used to provide the noisy dataset. FARSDAT is a Persian read speech composed of 6080 phonetically balanced sentences, 20 sentences uttered by each of 304 speakers. All sentences were randomly partitioned into three sets, namely training, development, and test, that respectively consist of about 75%, 10% and 15% of the whole corresponding corpus. There is no overlap between the speakers of these three sets.

We considered noise data from different corpora such as NOISEX-92 corpus [28], Demand [29] and noise free sounds [30]. In order to prepare the noisy dataset, all clean utterances from the either training or development set were corrupted with different noise types at seven SNRs from the set {-10, -5, 0, 5, 10, 15, 20} dB. By focusing on a main challenge in speech enhancement, i.e. mismatched noise conditions, we considered different noise types and SNRs for generating noisy speech utterances in the test set. To this end, 100 clean sentences were randomly selected from the test set, and each was corrupted with five unseen noise types from the set {*Leopard*, *Destroyer engine*, *Restaurant*, *Kitchen*, and *HF channel*} and six SNRs from the set {-7.5, -2.5, 2.5, 7.5, 12.5, 17.5} dB. Accordingly, 3000 noisy utterances were provided to evaluate the performance of enhancement methods.

We evaluated the performance of the enhancement methods based on two objective measures representing both quality and intelligibility of enhanced speech. The Perceptual Evaluation of Speech Quality (PESQ) [31] is considered as a speech quality measure that adopts the values in the intervals of [-0.5,4.5]. Moreover, the Extended Short-Time Objective Intelligibility (ESTOI) [32] is related to the intelligibility and limited to 0 to 1. For both measures, higher values are better.

#### B. Results and Discussion

For performance comparison, we considered four methods as below:

- OMLSA: A well-known statistical-based spectral amplitude estimator presented in [4] for speech enhancement,

- UNet: A U-Net architecture trained with conventional magnitude spectrum images represented in linear frequency and logarithm amplitude domains.

- UNet-MelPow: A U-Net architecture trained with perceptually-modified magnitude spectrum images represented in Mel frequency and power-law amplitude domains.

- Proposed (VGG19-UNet-MelPow): A U-Net architecture with VGG19 embedded at its encoder part and trained with perceptually-modified magnitude spectrum images represented in Mel frequency and power-law amplitude domains.

Tables I and II demonstrate the comparisons of these methods on the test set using PESQ and ESTOI measures, respectively. The results for the noisy case were also shown. In these tables, the average performances for different SNR levels, ranging from a very low SNR (-7.5 dB) to a high one (17.5 dB), were computed. Finally, the total average on all six SNRs for each approach was also reported.

Regarding Tables I and II, the following results can be expressed:

- All three methods employed deep neural networks significantly outperformed the statistical-based enhancer, i.e. OMLSA, based on both quality and intelligibility measures. According to the speech quality measure, i.e. PESQ, more improvements via deep learning methods compared to the OMLSA were obtained at more SNR levels. However, given the speech intelligibility index, i.e. ESTOI, more performance gains were achieved at lower SNRs.

- Comparing UNet and UNet-MelPow approaches, the latter yielded the better performances in almost all SNRs, especially in terms of the ESTOI measure. This observation represents that using power-law non-linearity for spectrum amplitude and Mel warping for spectrum frequency has a positive effect on speech enhancement measures, specifically speech intelligibility, compared to employing logarithm amplitude and linear frequency.
- The proposed approach shows the best performances for all six SNRs and both speech enhancement measures. For example, the average PESQ was improved from 2.31 to 2.49. This experiment verifies that using a very deep convolutional network such as VGG19 at the encoder part of the encoder-decoder structure as the U-Net considerably improves the speech enhancement performances. In fact, in the proposed image translation network, VGG19 plays the role of a very deep feature extractor of the noisy spectrum images.

Figures 2 and 3 demonstrate the average PESQ and ESTOI results of different methods over five unseen noise environments of the test set, respectively. Regarding both measures, it can be found that the worst performances of all methods were related to two difficult cases, namely *Restaurant* and *HF channel*, respectively. Similar to the previous findings, the UNet-MelPow approach was shown to be superior to the UNet one, for all five noise types, especially regarding the speech intelligibility measure, i.e. ESTOI.

Table I. Comparisons of methods based on PESQ measure

| Method | SNR (dB) | | | | | | |
|---|---|---|---|---|---|---|---|
| | -7.5 | -2.5 | 2.5 | 7.5 | 12.5 | 17.5 | Avg. |
| Noisy | 1.05 | 1.07 | 1.12 | 1.23 | 1.5 | 1.97 | 1.32 |
| OMLSA | 1.1 | 1.21 | 1.42 | 1.75 | 2.2 | 2.68 | 1.73 |
| UNet | 1.32 | 1.6 | 2 | 2.49 | 2.97 | 3.35 | 2.29 |
| UNet-MelPow | 1.34 | 1.63 | 2.04 | 2.52 | 2.97 | 3.34 | 2.31 |
| Proposed | 1.42 | 1.76 | 2.23 | 2.75 | 3.21 | 3.58 | **2.49** |

Table II. Comparisons of methods based on ESTOI measure

| Method | SNR (dB) | | | | | | |
|---|---|---|---|---|---|---|---|
| | -7.5 | -2.5 | 2.5 | 7.5 | 12.5 | 17.5 | Avg. |
| Noisy | 0.35 | 0.48 | 0.61 | 0.73 | 0.83 | 0.9 | 0.65 |
| OMLSA | 0.35 | 0.52 | 0.67 | 0.79 | 0.88 | 0.93 | 0.69 |
| UNet | 0.52 | 0.67 | 0.78 | 0.85 | 0.9 | 0.92 | 0.77 |
| UNet-MelPow | 0.54 | 0.69 | 0.79 | 0.86 | 0.91 | 0.94 | 0.79 |
| Proposed | 0.57 | 0.72 | 0.82 | 0.88 | 0.92 | 0.95 | **0.81** |

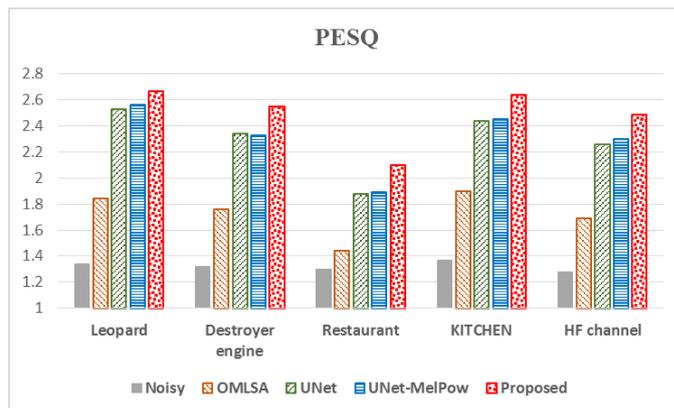

Figure 2. The average of PESQ values of different methods over five noise environments used in the test set.

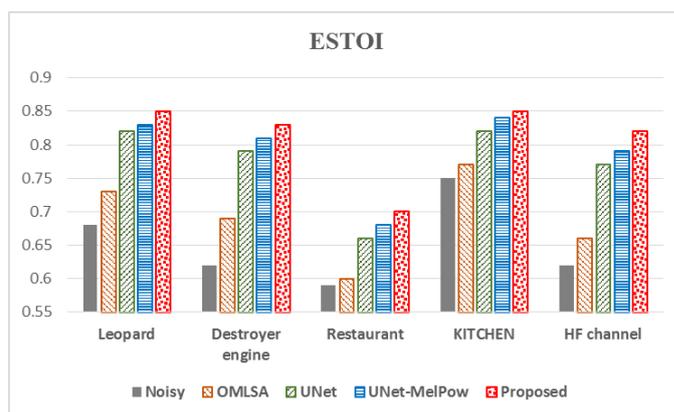

Figure 3. The average of ESTOI values of different methods over five noise environments used in the test set.

The enhanced spectrograms from noisy speech signal corrupted by *destroyer engine* noise at SNR = -2.5 dB using different methods were demonstrated in Figure 4. First, the OMLSA method (Figure 4(b)) has a little effect on reducing non-stationary noise components. By comparing Figures 4(c) and 4(d) to Figure 4(b), we can observe that using deep learning methods relative to the statistical-based enhancer (i.e. OMLSA) could desirably decrease noise components, and at the same time, keep well the formant and harmonic structures. Additionally, the proposed VGG19-UNet (Figure 4(d)) compared to the conventional UNet (Figure 4(c)) has obviously amplified the energy at main harmonics and generated the much more smooth harmonic structure (for example, consider two oval regions shown in Figure 4(d)).

It should be mentioned that for either UNet or UNet-MelPow approach, there are 512 filters at the bottleneck of the U-Net architecture consisting of about 7.7M trainable parameters. In contrast to, the number of trainable parameters for the proposed architecture, i.e. VGG19-UNet, is about 31M. Given this fact, we investigated the effect of increasing the number of parameters of the U-Net architecture, used in UNet-MelPow approach, on the speech enhancement performances.

parameters in the UNet-MelPow-1024 approach compared to UNet-MelPow, improved PESQ from 2.31 to 2.38 and the ESTOI from 0.79 to 0.795.

Table III. Comparisons of two methods UNet-MelPow-1024 with the proposed based on PESQ and ESTOI measures.

| Method | SNR (dB) | | | | | | |
|---|---|---|---|---|---|---|---|
| | -7.5 | -2.5 | 2.5 | 7.5 | 12.5 | 17.5 | Avg. |
| | PESQ | | | | | | |
| UNet-MelPow-1024 | 1.36 | 1.68 | 2.11 | 2.61 | 3.07 | 3.44 | 2.38 |
| Proposed | 1.42 | 1.76 | 2.23 | 2.75 | 3.21 | 3.58 | **2.49** |
| | ESTOI | | | | | | |
| UNet-MelPow-1024 | 0.55 | 0.70 | 0.80 | 0.87 | 0.91 | 0.94 | 0.795 |
| Proposed | 0.57 | 0.72 | 0.82 | 0.88 | 0.92 | 0.95 | **0.81** |

IV. CONCLUSION

Motivated by the importance of high speech quality and intelligibility for hearing aid and cochlear implant users under noisy listening conditions, we proposed a novel speech enhancement method based on deep neural networks. Accordingly, we suggested a novel image-to-image translation network, called VGG10-UNet, by embedding a deep FCN known as VGG19 at the encoder section of the conventional U-Net. The proposed network was used to translate the perceptually-modified spectrum images of noisy signals to the correspond images of clean ones. The perceptually-modified spectrum images are generated in Mel frequency and power-law non-linearity amplitude domains, instead of conventionally linear frequency and logarithm amplitude domains, respectively. It was shown that this modification on input spectrum features has a positive effect on speech enhancement measures, especially speech intelligibility. Finally, we observed that using a very deep feature extractor such as VGG19 at the encoder of an image translation network considerably improves both speech quality and intelligibility measures under different mismatched conditions represented by unseen noise types and SNR levels. Further improvements by employing different inputs and targets to the novel deep architectures can be considered as a future work.

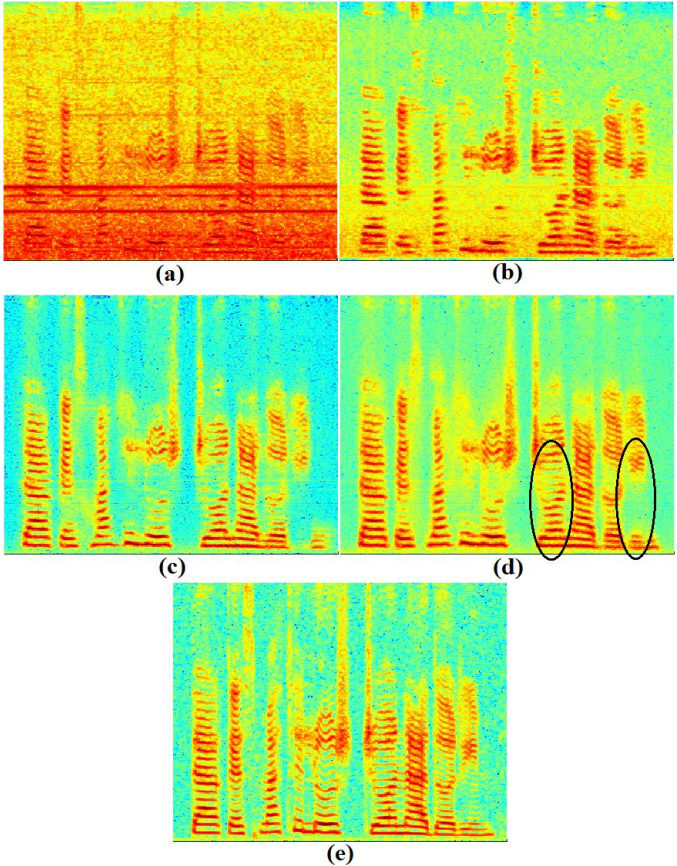

Figure 4. Spectrograms of a speech corrupted with *Destroyer engine* noise at SNR = -2.5 dB. (a) noisy speech (PESQ=1.06, ESTOI=0.46), (b) OMLSA (PESQ=1.37, ESTOI=0.64), (c) UNet-MelPow (PESQ=1.63, ESTOI=0.75), (d) Proposed: VGG19-UNet-MelPow (PESQ=2.07, ESTOI=0.80), and (e) the clean speech (PESQ=4.5, ESTOI=1.0).

To this end, we doubled the number of filters in all convolutional layers of the U-Net such that there were 1024 filters at its bottleneck. This change results in about 31M parameters for the U-Net architecture that are equal to the number of the parameters in the proposed one. In the following table, we called the UNet-MelPow approach using U-Net with 1024 filters at its bottleneck as UNet-MelPow-1024 and compared its performances with the proposed approach.

Given Table III, it can be observed that despite the same number of trainable parameters in both methods, the proposed approach still outperforms the UNet-MelPow-1024 regarding both PESQ and ESTOI measures. This could be due to the deeper structure of the encoder (or equivalently deeper feature extractor) employed in the proposed VGG19-UNet architecture compared to that of the U-Net in the UNet-MelPow-1024 method. Notice that there are 16 convolutional layers at the encoder of the proposed VGG19-UNet as opposed to only 10 convolutional layers for that of the conventional U-Net. As expected, however, increasing the number of trainable

REFERENCE

[1] K. Hermus, P. Wambacq, and H. Van, "A Review of Signal Subspace Speech Enhancement and Its Application to Noise Robust Speech Recognition," *EURASIP J. Adv. Signal Process.*, vol. 2007, no. 1, p. 045821, 2006.
[2] A. El-Solh, A. Cuhadar, and R. A. Goubran, "Evaluation of Speech Enhancement Techniques for Speaker Identification in Noisy Environments," in *Ninth IEEE International Symposium on Multimedia Workshops (ISMW 2007)*, 2007, pp. 235–239.
[3] Y. H. Lai, F. Chen, S. S. Wang, X. Lu, Y. Tsao, and C. H. Lee, "A Deep Denoising Autoencoder Approach to Improving the Intelligibility

of Vocoded Speech in Cochlear Implant Simulation," *IEEE Trans. Biomed. Eng.*, vol. 64, no. 7, pp. 1568–1578, 2017.

[4] I. Cohen and B. Berdugo, "Speech enhancement for non-stationary noise environments," *Signal Processing*, vol. 81, no. 11, pp. 2403–2418, Nov. 2001.

[5] X. Lu, S. Matsuda, C. Hori, and H. Kashioka, "Speech restoration based on deep learning autoencoder with layer-wised pretraining," *Interspeech*, pp. 1504–1507, 2012.

[6] X. Lu, Y. Tsao, S. Matsuda, and C. Hori, "Speech enhancement based on deep denoising and Auto-Encoder and Article," *Interspeech*, no. August, pp. 436–440, 2013.

[7] Y. Xu, J. Du, L.-R. Dai, and C.-H. Lee, "An Experimental Study on Speech Enhancement BasedonDeepNeuralNetworks," *IEEE Signal Process. Lett.*, vol. 21, no. 1, pp. 65–68, 2014.

[8] Y. Xu, J. Du, L. R. Dai, and C. H. Lee, "A regression approach to speech enhancement based on deep neural networks," *IEEE/ACM Trans. Audio Speech Lang. Process.*, vol. 23, no. 1, pp. 7–19, 2015.

[9] K. Han, Y. Wang, D. L. Wang, W. S. Woods, I. Merks, and T. Zhang, "Learning Spectral Mapping for Speech Dereverberation and Denoising," *IEEE Trans. Audio, Speech Lang. Process.*, vol. 23, no. 6, pp. 982–992, 2015.

[10] A. Kumar and D. Florencio, "Speech enhancement in multiple-noise conditions using deep neural networks," *Proc. Annu. Conf. Int. Speech Commun. Assoc. INTERSPEECH*, vol. 08-12-Sept, pp. 3738–3742, 2016.

[11] T. G. Kang, J. W. Shin, and N. S. Kim, "DNN-based monaural speech enhancement with temporal and spectral variations equalization," *Digit. Signal Process. A Rev. J.*, vol. 74, pp. 102–110, 2018.

[12] J. M. Martin-Donas, A. M. Gomez, J. A. Gonzalez, and A. M. Peinado, "A Deep Learning Loss Function Based on the Perceptual Evaluation of the Speech Quality," *IEEE Signal Process. Lett.*, vol. 25, no. 11, pp. 1680–1684, 2018.

[13] S. R. Park and J. W. Lee, "A Fully Convolutional Neural Network for Speech Enhancement," in *Proc. INTERSPEECH*, 2016, pp. 1993–1997.

[14] A. Jodeiri *et al.*, "Estimation of Pelvic Sagital Inclanation from Anteroposterior Radiograph Using Convolutional Neural Networks: Proof-of-Concept Study," in *EPiC Series in Health Sciences 2*, 2018, pp. 114–118.

[15] A. Pandey and D. Wang, "A New Framework for Supervised Speech Enhancement in the Time Domain," in *Proceedings of INTERSPEECH*, 2018, no. September, pp. 1136–1140.

[16] S.-W. Fu, Y. Tsao, X. Lu, and H. Kawai, "Raw waveform-based speech enhancement by fully convolutional networks," in *2017 Asia-Pacific Signal and Information Processing Association Annual Summit and Conference (APSIPA ASC)*, 2017, pp. 006–012.

[17] K. Tan and D. Wang, "A Convolutional Recurrent Neural Network for Real-Time Speech Enhancement," in *Proceedings of Interspeech*, 2018, pp. 3229–3233.

[18] K. Tan, X. Zhang, and D. Wang, "Real-time Speech Enhancement Using an Efficient Convolutional Recurrent Network for Dual-microphone Mobile Phones in Close-talk Scenarios," in *ICASSP 2019 - 2019 IEEE International Conference on Acoustics, Speech and Signal Processing (ICASSP)*, 2019, pp. 5751–5755.

[19] W. Han, C. Wu, X. Zhang, M. Sun, and G. Min, "Speech enhancement based on improved deep neural networks with MMSE pretreatment features," in *2016 IEEE 13th International Conference on Signal Processing (ICSP)*, 2016, pp. 1140–1145.

[20] A. Nicolson and K. K. Paliwal, "Deep learning for minimum mean-square error approaches to speech enhancement," *Speech Commun.*, vol. 111, pp. 44–55, 2019.

[21] S. Pascual, A. Bonafonte, and J. Serra, "SEGAN: Speech enhancement generative adversarial network," in *Proceedings of INTERSPEECH*, 2017, pp. 3642–3646.

[22] O. Ronneberger, P. Fischer, and T. Brox, "U-net: Convolutional networks for biomedical image segmentation," *Lect. Notes Comput. Sci. (including Subser. Lect. Notes Artif. Intell. Lect. Notes Bioinformatics)*, vol. 9351, pp. 234–241, 2015.

[23] K. Simonyan and A. Zisserman, "Very Deep Convolutional Networks for Large-Scale Image Recognition," in *arXiv preprint arXiv:1409.1556*, 2014.

[24] H. B. Kashani, A. Jodeiri, M. M. Goodarzi, and S. Gholamdokht Firooz, "Image to Image Translation based on Convolutional Neural Network Approach for Speech Declipping," in *4th Conference on Technology In Electrical and Computer Engineering (ETECH 2019), arXiv:1910.12116*, 2019.

[25] H. B. Kashani, A. Sayadiyan, and H. Sheikhzadeh, "Vowel detection using a perceptually-enhanced spectrum matching conditioned to phonetic context and speaker identity," *Speech Commun.*, vol. 91, pp. 28–48, 2017.

[26] H. B. Kashani and A. Sayadiyan, "Sequential use of spectral models to reduce deletion and insertion errors in vowel detection," *Comput. Speech Lang.*, vol. 50, pp. 105–125, 2018.

[27] M. Bijankhan, J. Sheikhzadegan, M. R. Roohani, Y. Samareh, C. Lucas, and M. Tebyani, "FARSDAT-The speech database of Farsi spoken language," in *Proceedings of the Australian Conference on Speech Science and Technology*, 1994, vol. 2, no. 0, pp. 826–831.

[28] A. Varga and H. J. M. Steeneken, "Assessment for automatic speech recognition: II. NOISEX-92: A database and an experiment to study the effect of additive noise on speech recognition systems," *Speech Commun.*, vol. 12, no. 3, pp. 247–251, 1993.

[29] J. Thiemann, N. Ito, and E. Vincent, "Diverse Environments Multichannel Acoustic Noise Database (DEMAND)," 2013. [Online]. Available: http://parole.loria.fr/DEMAND/.

[30] "FreeSound," 2015. [Online]. Available: https://freesound.org/.

[31] A. W. Rix, J. G. Beerends, M. P. Hollier, and A. P. Hekstra, "Perceptual evaluation of speech quality (PESQ)-a new method for speech quality assessment of telephone networks and codecs," in *2001 IEEE International Conference on Acoustics, Speech, and Signal Processing. Proceedings (Cat. No.01CH37221)*, 2001, pp. 749–752.

[32] J. Jensen and C. H. Taal, "An algorithm for predicting the intelligibility of speech masked by modulated noise maskers," *IEEE/ACM Trans. Audio, Speech, Lang. Process.*, vol. 24, no. 11, pp. 2009–2022, 2016.